\documentclass[aps,prl,reprint,superscriptaddress]{revtex4-2} 
\usepackage[dvipsnames]{xcolor}
\usepackage[colorlinks=true,urlcolor=blue,citecolor=purple,linkcolor=red]{hyperref}
\usepackage{amssymb,amsmath,amsfonts}
\usepackage{epsfig}
\usepackage{graphicx}
\usepackage{epstopdf}
\usepackage{natbib}
\usepackage{braket}

\def\clock{{\count0=\time
		\divide\count0 60
		\ifnum\count0<10 0\fi\the\count0
		\multiply\count0 -60 \advance\count0 \time
		:\ifnum\count0<10 0\fi \the\count0
}}
\newcommand{\timestamp}{{\small\vbox{\hbox{\tt\jobname.tex}
			\hbox{\the\day/\the\month/\the\year, \clock}}}}

\hypersetup{%
	pdftitle   = {Boundary Energy-Momentum Tensors for Asymptotically Flat Spacetimes},
	pdfkeywords = {Flat space holography, Carrollian geometry, holographic renormalisation, holographic reconstruction, boundary energy-momentum tensor},
	pdfauthor  = {Jelle Hartong, Emil Have, Vijay Nenmeli, Gerben Oling},
}

\makeatletter
\g@addto@macro\bfseries{\boldmath}
\makeatother

\newcommand{\C}{\mathbb{C}}

\newcommand{\nn}{\nonumber}

\newcommand{\be}{\begin{eqnarray}}
	\newcommand{\ee}{\end{eqnarray}}
\newcommand{\beq}{\begin{eqnarray}}
	\newcommand{\eeq}{\end{eqnarray}}

\newcommand{\beqa}{\begin{eqnarray}}
	\newcommand{\eeqa}{\end{eqnarray}}

\usepackage{mathrsfs}

\newcommand{\SO}{\operatorname{SO}}

\newcommand{\scri}{\mathcal{I}}

\ifdefined\C
\renewcommand{\C}{\boldsymbol{C}}
\else
\newcommand{\C}{\boldsymbol{C}}
\fi

\definecolor{gris}{rgb}{0.5,0.5,0.5}
\definecolor{darkgreen}{rgb}{0.0,0.5,0.0}

\allowdisplaybreaks[0]
\begin{document}

	\title{Boundary Energy-Momentum Tensors for Asymptotically Flat Spacetimes}
	
	\author{Jelle Hartong}
	\email{jelle.hartong@ed.ac.uk}
	\affiliation{School of Mathematics and Maxwell Institute for Mathematical Sciences,\\
	University of Edinburgh, Peter Guthrie Tait road, Edinburgh EH9 3FD, UK}

	\author{Emil Have}
	\email{emil.have@nbi.ku.dk}
	\affiliation{Center of Gravity, Niels Bohr Institute, University of Copenhagen, Blegdamsvej 17, DK-2100 Copenhagen Ø, Denmark}
	
	\author{Vijay Nenmeli}
	\email{v.v.nenmeli@sms.ed.ac.uk}
	\affiliation{School of Mathematics and Maxwell Institute for Mathematical Sciences,\\
	University of Edinburgh, Peter Guthrie Tait road, Edinburgh EH9 3FD, UK}
	
	\author{Gerben Oling}
	\email{gerben.oling@ed.ac.uk}
	\affiliation{School of Mathematics and Maxwell Institute for Mathematical Sciences,\\
	University of Edinburgh, Peter Guthrie Tait road, Edinburgh EH9 3FD, UK}
	
	\begin{abstract}
We consider 3D and 4D asymptotically flat spacetimes near future null infinity endowed with the most general allowed Carroll geometry. We define a boundary energy-momentum tensor by varying the on-shell action with respect to the Carroll metric data. This requires adding counterterms to the Einstein--Hilbert action. We show that, in 4D, the shear is on par with the Carroll metric data. Their combined response defines a boundary energy-momentum-news complex whose diffeomorphism Ward identity is equivalent to the Bondi mass and angular momentum loss equations. Weyl invariance leads to an identity for the trace of the energy-momentum tensor, and local Carroll boosts are anomalous in 3D and in 4D. 
	\end{abstract}

\maketitle
\textbf{Introduction.} 
Understanding radiation and its relation to the asymptotic structure of spacetime
is one of the fundamental questions in gravitational physics.
The groundbreaking analysis of Bondi, van der Burg, Metzner and Sachs~\cite{Bondi:1962px,Sachs:1962wk,Mädler:2016} revealed 
an infinite-dimensional symmetry algebra for asymptotically flat spacetimes. This BMS algebra has played a key role in modern developments involving asymptotic symmetries and soft theorems~\cite{Strominger:2013jfa,Strominger:2017zoo}
and the celestial holography programme~\cite{Pasterski:2016qvg,Pasterski:2021raf}.
The time evolution of the BMS charges is dictated by the shear near future null infinity $\scri^+$. This time evolution stems from the Bondi loss equations, which govern the evolution of the Bondi mass and angular momentum aspect.

An important question is whether the Bondi loss equations and BMS charges can be formulated in terms of a boundary energy-momentum tensor (EMT). In this work, we show that the answer to this is affirmative, with more details to be provided in upcoming work~\cite{Hartong:2025WIP}. A crucial insight is that the shear should be viewed as an integral part of the boundary geometry, on the same footing as the boundary metric data. For earlier and related work addressing this problem, see~\cite{Chandrasekaran:2021hxc,Adami:2024rkr,Freidel:2024emv,Riello:2024uvs,Bhambure:2024ftz,Ciambelli:2025mex}.

This work rests on the realisation that null infinity is a (conformal) Carrollian manifold \cite{Duval:2014uva} and that one can use bulk vielbeine to endow it with an arbitrary Carrollian geometry \cite{Hartong:2015usd}.
The latter paper, which focused on the 3-dimensional case, is the starting point of this work~\footnote{The main difference between this work and~\cite{Hartong:2015usd} relates to the choice of cut-off surface which in \cite{Hartong:2015usd} was required to be null whereas here it is taken to be an $r=\text{cst}$ hypersurface where $r$ is the Bondi radial coordinate.}.
We construct the asymptotic solution space (in the absence of logs) near $\scri^+$ 
for 3- and 4-dimensional bulk spacetimes, where we allow the boundary to be described by the most general Carrollian geometry that is allowed by the Einstein equations. It turns out that the Einstein equations impose one constraint on the boundary geometry in 4D. We do not fix coordinates on the boundary, and we also do not use a $2+1$ split of the boundary geometry (or a $1+1$ split for the case of a 3D bulk).
Instead, we use a gauge choice that generalises the usual Bondi--Sachs gauge to what we call Carroll-covariant Bondi--Sachs gauge. 
Our work generalises previous works that aimed at enlarging the solution space near future null infinity~\cite{Barnich:2010eb,Campiglia:2015yka,Geiller:2022vto,Donnay:2023mrd}.
It also connects with a series of works discussing the existence of nonzero Carroll energy flux in holographic settings~\cite{Ciambelli:2018wre,Mittal:2022ywl,Campoleoni:2023fug}.

By varying and renormalising the on-shell action, we then define an EMT-news complex as the responses to varying the Carrollian boundary metric data and the shear. This is done in the spirit of what in the AdS/CFT context has come to be known as holographic renormalisation \cite{Henningson:1998gx,Balasubramanian:1999re,deHaro:2000vlm,Skenderis:2000in,Bianchi:2001kw}. This allows us to show that the Bondi loss equations are diffeomorphism Ward identities for the EMT-news complex. Likewise, we obtain Ward identities for boundary Weyl transformations and Carroll boosts, and find that the latter are anomalous in 3D and 4D.

It should also be possible to achieve our results using covariant phase space methods~\cite{Compere:2018ylh,McNees:2023tus,Capone:2023roc,Geiller:2024amx,McNees:2024iyu,Riello:2024uvs}. In this regard we especially expect there to be overlap with the results of~\cite{Riello:2024uvs}.
Finally, we expect these results to be relevant in the search for a dual Carrollian field theory description of bulk gravitational degrees of freedom~\cite{Donnay:2022aba,Bagchi:2022emh,Donnay:2022wvx}.

\textbf{Penrose construction for the null boundary.} 
Consider a $(d+2)$-dimensional Lorentzian geometry $(\mathcal{M},g)$ with line element
\begin{equation}
\label{eq:line-element}
    ds^2 = -2UV+E^a E^a\,,
\end{equation}
where $a=1,\ldots,d$ are spatial tangent space indices, and where $U, V$ are null and $E^a$ are spatial 1-forms.
The line element~\eqref{eq:line-element} is invariant under local $\SO(d+1,1)$ transformations, which split into $SO(d)$ rotations, null boosts ($\Lambda$), and two sets of null rotations $(\Lambda^a,\Sigma^a)$ with axes $U$ and $V$, respectively. Infinitesimally these are
\begin{subequations}
    \begin{eqnarray}
    \delta E^a & = & \Lambda^{ab} E^b+\Lambda^a U+\Sigma^a V\,,\\
    \delta U & = & \Sigma^a E^a+\Lambda U\,,\quad
    \delta V  =  \Lambda^a E^a-\Lambda V\,,
\end{eqnarray}
\end{subequations}
where $\Lambda^{ab}$ is antisymmetric.

Let $\bar{\mathcal{M}}$ be a $(d+2)$-dimensional manifold with boundary $\partial\bar{\mathcal{M}}$ and interior $\mathcal{M}$. Let there be a regular null-bein basis $\bar U$, $\bar V$ and $\bar E^a$ on $\bar{\mathcal{M}}$. 
Let $\Omega$ be a function (the defining function~\cite{Penrose:1962ij}) that is zero on $\partial\bar{\mathcal{M}}$, positive on the interior $\mathcal{M}$, and such that $d\Omega$ is nonzero on the boundary $\partial\bar{\mathcal{M}}$. We will require that $d\Omega$ has vanishing norm on $\partial\bar{\mathcal{M}}$. We define the 1-forms $\bar U$, $\bar V$ and $\bar E^a$ in terms of $\Omega$ and the 1-forms $U$, $V$ and $E^a$ as follows
\begin{equation}
    \bar V=V\,,\qquad\bar U=\Omega^{2} U\,,\qquad  \bar E^a=\Omega E^a\,,
\end{equation}
so that $d\bar s^2=-2\bar U\bar V+\bar E^a \bar E^a=\Omega^2 ds^2$.
We require
\begin{eqnarray}
 && \hspace{-.5cm}  \bar V\cdot\partial\Omega\big|_{\Omega=0}\neq 0\,,~ \bar U\cdot\partial\Omega\big|_{\Omega=0}= 0\,,~ \bar E_a\cdot\partial\Omega\big|_{\Omega=0}= 0\,,\label{eq:vectorbdry}\\
&&   \hspace{-.5cm}  \bar V\cdot \partial F\big|_{\Omega=0}=\mathcal{O}(1)\,,~ \bar U\cdot \partial F\big|_{\Omega=0}\neq 0\,,~ \bar E_a\cdot\partial F\big|_{\Omega=0}\neq 0\,,\nn
\end{eqnarray}
where we evaluate derivatives of $\Omega$ before setting $\Omega=0$ and where $F$ is any scalar function on the boundary. Any two defining functions $\Omega$ and $\Omega'$ related by $\Omega'=e^f\Omega$, with $f$ a scalar on the boundary, are equally good choices. This leads to a conformal class of boundary metric data.

We will denote bulk coordinates as $x^M=(r, x^\mu)$ where $x^\mu$ are coordinates on $\partial\bar{\mathcal{M}}$ (henceforth denoted as $\scri^+$). We can take $\Omega=r^{-1}$. The coordinate $r$ will be a radial coordinate in terms of which we expand the near-boundary metric.

It will be convenient to partially fix local Lorentz transformations and diffeomorphisms by setting $V^M=-(\partial_r)^M$. Equation \eqref{eq:vectorbdry} 
tells us that
\begin{equation}
    \begin{split}
& U^r = \mathcal{O}(r)\,,\qquad U^\mu=v^\mu+\mathcal{O}(r^{-1})\,,\\
& E^r_a=\mathcal{O}(1)\,,\qquad E^\mu_a=r^{-1}e^\mu_a+\mathcal{O}(r^{-2})\,,
\end{split}
\end{equation}
where $v^\mu$ and $e^\mu_a$ form a complete set of nowhere vanishing boundary frame fields. We assumed that the expansion in $r$ is in powers of $r^{-1}$, although this may be modified by allowing for a polylogarithmic expansion (see~\cite{Hartong:2025WIP} for more details on log terms). This implies that 
\begin{equation}
\label{eq:Falloff}
\begin{split}
     E^a_\mu &= re^a_\mu+
    \mathcal{O}(1)\,,\quad V_\mu = \tau_\mu+\mathcal{O}(r^{-1}) \,,\\   U_{\mu} &= rb_\mu+\mathcal{O}(1)\,,\quad
     U_r = 1\,,\quad    E_r^a = 0\,,\quad
    V_r = 0\,,
\end{split}
\end{equation}
where $(\tau_\mu\,,e_\mu^a)$ is the inverse of $(-v^\mu\,,e^\mu_a)$.

\textbf{Carroll-covariant Bondi--Sachs gauge.}
To see that the geometry on the Penrose boundary is Carrollian,
we consider the asymptotic behaviour of bulk local Lorentz transformations and diffeomorphisms,
following~\cite{Hartong:2015usd}.
These respect the boundary conditions \eqref{eq:Falloff}, subject to the gauge-fixing condition $V^M=-(\partial_r)^M$ and $\delta V^M=\mathcal{L}_\xi V^M-\Lambda V^M+\Lambda^a E_a^M=0$, if and only if
\begin{eqnarray}
\label{eq:bulkdiffeos2}
            &&\Lambda^{ab} = \lambda^{ab}+\mathcal{O}(r^{-1})\,,\quad \Lambda = -\Lambda_D+\mathcal{O}(r^{-1})\,,\nn\\
     &&\Sigma^a = \sigma^a+\mathcal{O}(r^{-1})\,,~~~\,
    \Lambda^a = r^{-1}\lambda^a+\mathcal{O}(r^{-2})\,,\\
     &&\xi^r = r\Lambda_D+\mathcal{O}(1)\,,~~ \xi^\mu = \chi^\mu+r^{-1}\lambda^\mu+\mathcal{O}(r^{-2})\,,\nn
\end{eqnarray}
where $\lambda^\mu=\lambda^a e_a^\mu$. The action on the boundary data is 
\begin{subequations}
\label{eq:bdary-trafos}
    \begin{eqnarray}
\delta\tau_\mu & = & \mathcal{L}_\chi\tau_\mu+\Lambda_D\tau_\mu+\lambda^a e^a_\mu\,,\label{eq:totrafotau}\\
\delta h_{\mu\nu} & = & \mathcal{L}_\chi h_{\mu\nu}+2\Lambda_D h_{\mu\nu} \,,\label{eq:totrafoh}\\
\delta b_\mu & = & \mathcal{L}_\chi b_\mu+\partial_\mu\Lambda_D+\sigma^a e^a_\mu\,,\label{eq:btrafo}
\end{eqnarray}
\end{subequations}
where we defined $h_{\mu\nu}=\delta_{ab}e^a_\mu e^b_\nu$. From these transformations we infer that $b_\mu$ is a boundary Weyl connection for Weyl transformations with parameter $\Lambda_D$, while $\lambda^a$ is the Carroll boost parameter.

We fix the null rotations with parameter $\Sigma^a$ by setting $E^\mu_a U_\mu=0$,
which is equivalent to $U_\mu=\frac{1}{2}S V_\mu$ for some bulk scalar field $S = \mathcal{O}(r)$,
and we can further fix bulk diffeomorphisms by going to what we dub the Carroll-covariant Bondi--Sachs gauge by fixing the following components of the bulk Levi-Civita connection
\begin{equation}
\Gamma^\rho_{rr}=0\,,\qquad\Gamma^\rho_{\rho r}=dr^{-1}\,.     
\end{equation}
Dropping the gauge fixing condition $\Gamma^\rho_{\rho r}=dr^{-1}$ gives a Carroll-covariant version of the partial Bondi gauge~\cite{Geiller:2022vto}.
The condition $\Gamma^\rho_{rr}=0$ is equivalent to setting $V_\mu=e^\beta\tau_\mu$ for some bulk scalar field $\beta = \mathcal{O}(r^{-1})$. This implies that $V^M=-(\partial_r)^M$ is geodesic. In this gauge, the metric is 
\begin{subequations}\label{eq:metricdecom}
    \begin{eqnarray}
    &&ds^2 =-2e^\beta\tau_\mu dx^\mu dr+g_{\mu\nu}dx^\mu dx^\nu\,,\\ 
    &&g_{\mu\nu}=\Pi_{\mu\nu}-e^{2\beta}S\tau_\mu\tau_\nu\,,
\end{eqnarray}
\end{subequations}
so that the remaining bulk variables are $\beta, S$ and $\Pi_{\mu\nu}=\delta_{ab}E^a_\mu E^b_\nu$. The inverse metric is
\begin{equation}
    g^{rr}=S\,,\quad g^{\mu r}=U^\mu\,,\quad g^{\mu\nu}=\Pi^{\mu\nu}=\delta^{ab}E^\mu_a E^\nu_b\,.
\end{equation}
Note that $\Pi_{\mu\nu}$ has signature $(0,1,\cdots,1)$ because $U^\mu\Pi_{\mu\nu}=0$, allowing us to solve for $v^\mu v^\nu\Pi_{\mu\nu}$ order by order in $1/r$. The condition $\Gamma^\rho_{\rho r}=dr^{-1}$ allows us to solve for $h^{\mu\nu}\Pi_{\mu\nu}$ order by order in $r^{-1}$.
We use the Einstein equations to find $\beta$, $S$, $ h^\rho_\mu v^\sigma\Pi_{\rho\sigma}$ and $\mathcal{L}_v\left(h^\rho_{\langle\mu}h^\sigma_{\nu\rangle}\Pi_{\rho\sigma}\right)$ order by order in $r^{-1}$.
Here, we defined the boundary spatial projector $h^\mu_\nu = e^a_\nu e^\mu_a = \delta^\mu_\nu + v^\mu\tau_\nu$~\footnote{A boundary tensor $X^\mu{_\nu}$ is spatial iff $X^\mu{_\nu} = h^\mu_\rho h^\sigma_\nu X^\rho{_\sigma}$}, and the angle brackets denote the symmetric tracefree (STF) part.

\textbf{Solving Einstein's equations asymptotically.} 
From solving $R_{rr}=0$ at order $r^{-3}$, $R_{\mu r}=0$ at order $r^{-1}$, and  $R_{\mu\nu}=0$ at order $r$ we learn that 
\begin{subequations}
\begin{eqnarray}
\Pi_{\mu\nu}&=&r^2 h_{\mu\nu}+r\left(C_{\mu\nu}-2\tau_{(\mu}a_{\nu)}\right)+\mathcal{O}(1)\,,\\
S &=& \frac{2}{d}Kr+\mathcal{O}(1)\,,\quad  \beta = \mathcal{O}(r^{-2})\,,\\
K_{\mu\nu} & = & \frac{1}{d}K h_{\mu\nu}\,,\label{eq:K-constraint}
\end{eqnarray}
\end{subequations}
where $K_{\mu\nu}=-\frac{1}{2}\mathcal{L}_v h_{\mu\nu}$, $K=h^{\mu\nu}K_{\mu\nu}$, and $a_\mu=\mathcal{L}_v\tau_\mu$ with $\mathcal{L}_v$ the Lie derivative along $v^\mu$. Equation \eqref{eq:K-constraint} is a constraint in $d\ge 2$. The field $C_{\mu\nu}$ is the shear (spatial and STF, i.e., $v^\mu C_{\mu\nu}=h^{\mu\nu}C_{\mu\nu}=0$, and thus zero for $d=1$). It transforms as
\begin{equation}
\label{eq:shear-trafo}
    \hspace{-0.5em}\delta C_{\mu\nu} = \mathcal{L}_{\chi}C_{\mu\nu}+\Lambda_{D}C_{\mu\nu}+h^{\rho}_{\langle\mu}h^{\sigma}_{\nu\rangle}\left(\mathcal{L}_\lambda h_{\rho\sigma}+2a_\rho\lambda_\sigma\right)\,.
\end{equation}
The shear transforms under the same gauge transformations as $\tau_\mu$ and $h_{\mu\nu}$,
and we will later see that it forms a fundamental part of the boundary data.
In fact, all properties of the boundary metric and shear data, including the constraint~\eqref{eq:K-constraint}, can be obtained from the gauging of the conformal Carroll algebra \cite{Hartong:2025WIP2} (see \cite{Korovin:2017xqu,Herfray:2020rvq,Baulieu:2025itt,Fiorucci:2025twa} for similar statements and related work).

We briefly explain what happens at subsequent orders; for more details see \cite{Hartong:2025WIP}. The result depends on the dimension. For $d=1$ the expansion terminates, and we find
\begin{equation}\label{eq:3Dsol}
\begin{split}
        \Pi_{\mu\nu} & =  r^2 h_{\mu\nu}-2r\tau_{(\mu}a_{\nu)}+a^2\tau_\mu\tau_\nu-2\tau_{(\mu}P_{\nu)}\,,\\
    S & = 2rK+a^2+2M\,,\quad\beta=0\,,
\end{split}
\end{equation}
where $M$ and $P_\mu$ (with $v^\mu P_\mu=0$) correspond to the Bondi mass and angular momentum aspect.
They obey a conservation equation that we will come back to later.

We denote by $X^{(n)}$ the coefficient of $r^{-n}$ in the expansion of $X$. 
For $d=2$, the Einstein equations tell us  
\begin{eqnarray*}
   &&\hspace{-.3cm} \Pi^{(0)}_{\mu\nu} = a^2\tau_\mu\tau_\nu-2\tau_{(\mu}P^{(0)}_{\nu)}+D_{\mu\nu}+\frac{1}{2}F_{\mu\rho}C^\rho{}_\nu+\frac{1}{4}h_{\mu\nu}C^2 \,,\\
    &&\hspace{-.3cm}\Pi^{(1)}_{\mu\nu} =  \left(2a^\rho P^{(0)}_\rho-C^{\rho\sigma}a_\rho a_\sigma\right)\tau_\mu\tau_\nu-2\tau_{(\mu}P^{(1)}_{\nu)}+h^\rho_\mu h^\sigma_\nu\Pi^{(1)}_{\rho\sigma}\,,\\
    &&\hspace{-0.3cm}S^{(0)}  =  \frac{1}{2}\mathcal{R}+\frac{3}{2}e^{-1}\partial_\mu\left(e a^\mu\right)\,,\quad
    \beta^{(2)} = \frac{1}{16}\left(F^2-C^2\right)\,.
\end{eqnarray*}
For our purpose we do not need to know $h^\rho_\mu h^\sigma_\nu\Pi^{(1)}_{\rho\sigma}$.
In here we defined
\begin{equation}
    F_{\mu\nu} = h^\rho_\mu h^\sigma_\nu\left(\partial_\rho\tau_\sigma-\partial_\sigma\tau_\rho\right)\,,
\end{equation}
which measures the failure of $\tau_\mu$ to be hypersurface orthogonal.
We write $F^2=F^{\mu\nu}F_{\mu\nu}$ and similarly for $C^2$.
We raise and lower spatial indices using $h^{\mu\nu}$ and $h_{\mu\nu}$ and denote the measure on the boundary by $e=\text{det}\,\left(\tau_\mu\,,e^a_\mu\right)$. Furthermore, we defined
\begin{equation}
    P^{(0)}_\mu = -\frac{1}{2}\left(\mathcal{D}_\rho-2a_\rho\right)C^\rho{}_\mu+\frac{1}{2}\mathcal{D}_\rho F^\rho{}_\mu\,.
\end{equation}
The tensor $D_{\mu\nu}$ 
is spatial and STF with the properties
\begin{equation}
  \label{eq:D-tensor-properties}
    \mathcal{L}_v D_{\mu\nu}=0\,,\quad\left(\mathcal{D}_\rho-a_\rho\right)D^\rho{}_\nu=0\,.
\end{equation}
The latter property can be removed by adding a $r^{-1}\log r$ term to the expansion of $\Pi_{\mu\nu}$ (see~\cite{Barnich:2010eb,Geiller:2022vto}). 

In the above expressions and below we use the following symmetric affine connection~\footnote{It can be beneficial to add to this connection the term $\frac{1}{2}v^\rho C_{\mu\nu}$ because it gives the connection nicer transformation properties under Carroll boosts, i.e., we then find 
\begin{eqnarray*}
\hspace{-0.5cm}\delta_{\text{boost}}\mathcal{C}^\rho_{\mu\nu}  =  -\frac{1}{2}e^{-1}\partial_\sigma\left(e\lambda^\sigma\right) v^\rho h_{\mu\nu}+K v^\rho\tau_{(\mu}\lambda_{\nu)}\,.
\end{eqnarray*}
See \cite{Nguyen:2022zgs,Baulieu:2025itt} for related observations. However, for the purposes of this work such a term does not make much difference as it drops out of our expressions so we leave it out. We refer to~\cite{Hartong:2025WIP} for more background on the choice of connection.}
\begin{equation}
\label{eq:C-connection}
    \mathcal{C}^\rho_{\mu\nu}=-v^\rho\left(\partial_{(\mu}+a_{(\mu}\right)\tau_{\nu)}+\frac{1}{2}h^{\rho\sigma}\left(2\partial_{(\mu} h_{\nu)\sigma} - \partial_\sigma h_{\mu\nu}\right)\,.
\end{equation}
We denote the associated covariant derivative operator and curvature tensor by $\mathcal{D}_\mu$ and $\mathcal{R}_{\mu\nu\rho}{}^\sigma$, respectively.
In particular, we set $\mathcal{R}=h^{\mu\rho}\mathcal{R}_{\mu\nu\rho}{}^\nu$, which features in the expression for $S^{(0)}$ and which (for $d=2$) can be shown to be the only 2-derivative curvature scalar that cannot be written in terms of derivatives of $K$ and $a_\mu$.
We follow the curvature conventions of~\cite{Hansen:2020pqs}. This choice of connection is torsion-free and volume-form compatible (i.e., $\mathcal{C}^\rho_{\rho\nu}=e^{-1}\partial_\nu e$). It has the following action on the metric data
\begin{subequations}
\begin{eqnarray}
    \mathcal{D}_\mu\tau _\nu & = & \frac{1}{2}F_{\mu\nu}-\tau_\mu a_\nu\,,\\
    \mathcal{D}_\rho h_{\mu\nu} & = & -\frac{1}{2}K\left(\tau_\mu h_{\nu\rho}+\tau_\nu h_{\mu\rho}\right)\,.
\end{eqnarray}
\end{subequations}
Having zero torsion and $\mathcal{C}^\rho_{\rho\nu}=e^{-1}\partial_\nu e$ ensures that the curvature tensor satisfies the standard Bianchi identities and that the Ricci tensor is symmetric.
It also simplifies writing Lie derivatives in terms of covariant derivatives and integration by parts, as $\mathcal{D}_\mu X^\mu=e^{-1}\partial_\mu\left(e X^\mu\right)$.
The absence of torsion means that we cannot require the analogue of metric compatibility with respect to the Carrollian structure $(v^\mu,h_{\mu\nu})$.
Instead, we can demand that $\mathcal{D}_\rho h_{\mu\nu}$ and $\mathcal{D}_\rho v^\mu$ are proportional to $K_{\mu\nu}$, which is the unique 1-derivative Carroll boost-invariant tensor. Indeed, we have $\mathcal{D}_\mu v^\nu=-K_\mu{}^\nu=-\frac{1}{2}Kh_\mu^\nu$. 

For $d=2$, we cannot algebraically solve for $S^{(1)}$ and $P^{(1)}_\rho=h^\mu_\rho v^\nu\Pi^{(1)}_{\mu\nu}$ and these are related to the Bondi mass and angular momentum aspect in a manner that we will see below. They obey the Bondi mass and angular momentum loss equations. One of the main results of this work is that we can write those as diffeomorphism Ward identities or, in other words, as conservation equations of a boundary EMT-news complex (see equations \eqref{eq:diffeoWtimeproj} and \eqref{eq:diffeoWIspatialproj} below).

\textbf{Boundary variations and Ward identities.}
Before we consider the definition of a boundary energy-momentum tensor in GR let us first consider what to expect. To this end, consider some action functional $S[\tau,h,C]$ of the boundary data $\tau_\mu, h_{\mu\nu}$ and $ C_{\mu\nu}$. We will consider $d=1,2$ where for $d=1$ any spatial STF tensor is zero. Let us define the responses as
\begin{equation}
\label{eq:variationOS}
    \delta S = \int d^{d+1}x \,e\left(T^\mu\delta\tau_\mu+\frac{1}{2}T^{\mu\nu}\delta h_{\mu\nu}+\frac{1}{2} S^{\mu\nu}\delta C_{\mu\nu}\right)\,,
\end{equation}
where $T^\mu$ is the energy current and $T^{\mu\nu}$ the momentum-stress tensor
(see for example~\cite{deBoer:2020xlc}).
Because $C_{\mu\nu}$ is spatial and STF the variation of $C_{\mu\nu}$ is not fully arbitrary.
We can require $S^{\mu\nu}$ to be spatial and STF without loss of generality~\footnote{Anything of the form $\left(Xh^{\mu\nu}+Y^\mu v^\mu\right)\delta C_{\mu\nu}$ can be absorbed into $T^\mu$ and $T^{\mu\nu}$.}. As we will see, the response $S^{\mu\nu}$ to varying the shear is the news tensor.

When $d=2$, the boundary geometry is subject to the constraint~\eqref{eq:K-constraint} and so we need to vary within the space of solutions to this constraint~\footnote{This constraint cannot be lifted by adding log terms to the radial expansion.}. There are two equivalent ways of doing so. We can solve the constraint (fully covariantly) and the most general solution is~\cite{Hartong:2025WIP} 
\begin{equation}
    h_{\mu\nu}=H^2\left(\partial_\mu X\partial_\nu X+\partial_\mu Y\partial_\nu Y\right)\,,
\end{equation}
where $H$, $X$ and $Y$ are arbitrary scalar fields that can be varied freely. Alternatively, we can add a Lagrange multiplier term $S_{\text{LM}} = \tfrac{1}{2}\int d^{d+1}x e\chi^{\mu\nu}K_{\mu\nu}$ to the action functional, where $\chi^{\mu\nu}$ is a spatial and STF Lagrange multiplier.
In both cases, the result is that we cannot distinguish between $T^{\mu\nu}$ and $T^{\mu\nu}+t^{\mu\nu}$, where 
\begin{equation}\label{eq:t}
    t^{\mu\nu}=\frac{1}{2}h^{\mu\rho}h^{\nu\sigma}\mathcal{L}_v\chi_{\rho\sigma}-v^{(\mu}h^{\nu)\sigma}\left(\mathcal{D}_\rho-a_\rho\right)\chi^\rho{}_\sigma\,.
\end{equation}
When we demand that the action is invariant under diffeomorphisms, Weyl transformations and Carroll boosts (possibly up to an anomaly), the responses $T^\mu$, $T^{\mu\nu}$ and $S^{\mu\nu}$ obey various Ward identities. It can be shown that these Ward identities do not depend on $t^{\mu\nu}$, so it corresponds to an improvement transformation that preserves all Ward identities. It should be viewed as an ambiguity of the EMT.
If we require that \eqref{eq:variationOS} is invariant under \eqref{eq:totrafotau}, \eqref{eq:totrafoh} and \eqref{eq:shear-trafo} we obtain the following Ward identities
\begin{subequations}
\begin{align}
\begin{split}
0 &=  -e^{-1}\partial_\mu \left(e \left[T^\mu{}_\nu+S^{\mu\rho}C_{\rho\nu}\right]\right)+T^\mu\partial_\nu\tau_\mu\\
        &\quad\,+\frac{1}{2}T^{\mu\rho}\partial_\nu h_{\mu\rho}+\frac{1}{2}S^{\mu\rho}\partial_\nu C_{\mu\rho}\,,\label{eq:diffWI}
\end{split}
        \\
    0 & =  T^\mu\tau_\mu+T^{\mu\nu}h_{\mu\nu}+\frac{1}{2}S^{\mu\nu}C_{\mu\nu}\,,\label{eq:non-anomalous-Weyl-WI}\\
    0 & =  h_\sigma^\rho T^\sigma-\left(\mathcal{D}_\mu -a_\mu\right)S^{\mu\rho}\,,\label{eq:non-anomalous-boost-WI}
\end{align}
\end{subequations}
which are the diffeomorphism ($\chi^\mu$), Weyl ($\Lambda_D$) and Carroll boost ($\lambda_\mu=\lambda^a e^a_\mu$) Ward identities respectively. The combination $T^{\mu}{}_\nu=T^\mu\tau_\nu+T^{\mu\rho}h_{\rho\nu}$ is the EMT.

In a Carrollian theory without additional sources, Carroll boost invariance dictates that the energy flux $h_\sigma^\rho T^\sigma$ vanishes \cite{deBoer:2017ing,deBoer:2021jej}.
When we couple the theory to sources that transform under local Carroll boosts, such as the shear, this statement gets modified.
This was also recently emphasised in~\cite{Fiorucci:2025twa} (and see~\cite{Armas:2023dcz} for related discussions).
As we will see, in GR it is further modified as the Carroll boosts are anomalous.
The other residual symmetries in GR correspond to boundary diffeomorphisms and Weyl transformations and we will see that these are not anomalous.

When solving the Einstein equations in our Carroll-covariant Bondi--Sachs gauge~\eqref{eq:metricdecom}, in particular $U^\mu R_{\mu\nu} = 0$ at order $r^{-d}$, we encounter what are commonly referred to as the Bondi loss/conservation equations (for $d=2$/$d=1$). These can be written in the form \eqref{eq:diffWI} for a judicious choice of boundary EMT and news tensor. In~\cite{Hartong:2025WIP} we show explicitly how to do this by computing $U^\mu R_{\mu\nu} = 0$ at order $r^{-d}$ for the case of the most general boundary geometry.
In this Letter, we instead derive this result using ideas that are very reminiscent of holographic renormalisation familiar from AdS/CFT.

\textbf{A well-posed variational principle.}  
Varying the bulk Einstein--Hilbert action $S_{\text{EH}} = \int d^{d+2}x\,\sqrt{-g}R$ gives
\begin{equation*}
    \hspace{-0.3cm}\delta S_{\text{EH}}=\int d^{d+1} x\sqrt{-g}\,G_{MN}\delta g^{MN}+\int d^{d+1}x\, \partial_M\left(\sqrt{-g}J^M\right)\,,
\end{equation*}
where $G_{MN}$ is the Einstein tensor and $J^M=g^{NP}\delta \Gamma^M_{NP}-g^{MP}\delta\Gamma_{PN}^N$. The second term in the variation of $\delta S_{\text{EH}}$ leads to boundary integrals that depend on the region for which we demand there to be a well-posed variational principle. We focus on the term that arises for large values of $r$, and we consider a cut-off surface at $r=\Lambda$ for large $\Lambda$.
Focusing entirely on this term,
and using the variables in the metric decomposition~\eqref{eq:metricdecom},
we get 
\begin{equation}
  \delta S_\text{EH}
  = \cdots+\int_{r = \Lambda} d^{d+1}x\, E\, J^r\,,
\end{equation}
where we wrote $E:=\sqrt{-g}=\text{det}\,\left(V_\mu\,,E^a_\mu\right)=er^de^\beta$, and where, up to total derivatives,
\begin{eqnarray*}
    J^r &=& \delta\left(-E^{-1}\partial_\mu\left(EU^\mu\right)-dr^{-1}S-\partial_r S-2S\partial_r\beta\right) \\
    &&-\Gamma^r_{\mu\nu}\delta\Pi^{\mu\nu}-2\Gamma^r_{\mu r}\delta U^\mu+E^{-1}\partial_\mu\left(EU^\mu\right)E^{-1}\delta E\,.\nn
\end{eqnarray*}
Our goal will be to make sure that the variation of the EH action (plus appropriate boundary terms) near $\mathcal{I}^+$ is well-defined, in the sense that the variation of the total action in the large $r$ limit is of the form \eqref{eq:variationOS} on $\mathcal{I}^+$.

To make the variational principle well-posed, we need an extrinsic Gibbons--Hawking--York-type boundary term. However, our cut-off surface is neither spacelike nor timelike nor null. A GHY-type term on a generic surface was proposed in \cite{Parattu:2016trq}, which involves writing down a normal $1$-form $N_Mdx^M$ that is exact, along with a vector $V^M$ satisfying $V^M N_M=-1$. In our case, there are natural candidates for both $N_M$ and $V^M$, namely $N_M=\partial_M r$ and $V^M=-\delta^M_r$~\footnote{The counterterm depends on the choice of $V^M$. However, if we take a different $V'^M$ that also obeys $V'^M N_M=-1$, the counterterm changes by an intrinsic term involving tangential derivatives. Since at this stage the purpose is to fix the nature of the variational problem (i.e. one with Dirichlet boundary conditions) the choice of $V^M$ does not matter. Once we make a particular choice we need to check if any further intrinsic counterterms are needed.}. The extrinsic counterterm is then 
\begin{eqnarray}
    &&
    S_{\text{ext}}=a\int_{r = \Lambda} d^{d+1} x \sqrt{-g}\left(\delta^M_P+V^M N_P\right)\nabla_M N^P \nn\\
    &&
    = a\int_{r = \Lambda} d^{d+1} x \,E\left[E^{-1}\partial_\mu\left(E U^\mu\right)+dr^{-1}S+\frac{1}{2}\partial_r S+S\partial_r\beta\right]\nn\,,
\end{eqnarray}
where $a$ is a real constant. Taking $a=2$ removes radial derivatives of $\delta S$ and $\delta \beta$ in $J^r$ above, leading to
\begin{eqnarray}
&&\delta \left(S_{\text{EH}}+S_{\text{ext}}\right)
 = \cdots\\
 &&\hspace{-0.1cm} +\int_{r = \Lambda} d^{d+1}x\, E\left(\mathcal{T}^\mu\delta V_\mu+\frac{1}{2}\mathcal{T}^{\mu\nu}\delta\Pi_{\mu\nu}+dr^{-1}E^{-1}\delta\left(E S\right)\right)\,,\nn
\end{eqnarray}
where the last equality defines the objects $\mathcal{T}^\mu$ and $\mathcal{T}^{\mu\nu}$, which are precursors to the energy current and the momentum-stress tensor. Although we refrain from writing out their explicit expressions, we note that $\mathcal{T}^\mu = \mathcal{O}(r^{-1})$ and $\mathcal{T}^\mu = \mathcal{O}(r^{-3})$, which implies that the term $\mathcal{T}^\mu\delta V_\mu+\frac{1}{2}\mathcal{T}^{\mu\nu}\delta\Pi_{\mu\nu}$ is $\mathcal{O}(r^{-1})$. On the other hand, the the term $dr^{-1}E^{-1}\delta\left(E S\right)$ is $\mathcal{O}(1)$, so it dominates.
The latter term always leads to a variation of the free data $S^{(d-1)}$ that is not cancelled when we go to $\mathcal{O}(1)$. This is thus not of Dirichlet type. We can remove this by adding the second counterterm~\footnote{Factors of powers of $r$ in this and other counterterms can be understood by requiring invariance of the counterterm under bulk dilatations $r'=\lambda^{-1} r$ and $x'^\mu=\lambda x^\mu$.}
\begin{equation}
    S_{\text{norm}}=-d\int_{r=\Lambda}d^{d+1}x\, E r^{-1}S\,,
\end{equation}
which depends on the norm of the normal $N_M$
which is $g^{rr}=S$~\footnote{In the AdS setting, this term would just be a cosmological constant}. With this, we obtain 
\begin{equation}
\begin{split}
  &\delta \left(S_{\text{EH}}+S_{\text{ext}}+S_{\text{norm}}\right)
   =\label{eq:varSEH+Sext2}\\
   &\cdots+2\int_{r = \Lambda} d^{d+1}x\, E\left(\mathcal{T}^\mu\delta V_\mu+\frac{1}{2}\mathcal{T}^{\mu\nu}\delta\Pi_{\mu\nu}\right)\,.
\end{split}
\end{equation}
We now have a variational Dirichlet problem on the cut-off surface. Depending on the dimension, we may need to add additional ``intrinsic'' counterterms (which respect the nature of the variational problem) that remove divergences, while we are always free to add finite counterterms which lead to improvements. We will now go through this procedure for $d=1,2$ separately. 

\textbf{Boundary EMT for $d=1$.} 
Since $\beta = 0$ here, we get $E = re$, so the right-hand side of~\eqref{eq:varSEH+Sext2} is $\mathcal{O}(1)$.
To ensure the boundary EMT satisfies the Weyl Ward identity~\eqref{eq:non-anomalous-Weyl-WI} (which means it is traceless since there is no shear and news for $d=1$),
we add the finite counterterm
\begin{equation}
  \label{eq:3d-finite-counterterm-traceless}
    S_{\text{finite}}=-\int_{r = \Lambda} d^2x \,rE\,\Pi^{\mu\nu} A_\mu A_\nu\,,
\end{equation}
where we defined $A_\mu=\mathcal{L}_U V_\mu$.
This gives
\begin{equation*}
\begin{split}
  &\delta \left(S_{\text{EH}}+S_{\text{ext}}+S_{\text{norm}} + S_{\text{finite}}\right)\Big\vert_{r\to\infty}
   =\\
   &
   2\int d^{2}x\, e\left(T_{\text{ren}}^\mu\delta\tau_\mu + \frac{1}{2}T_{\text{ren}}^{\mu\nu}\delta h_{\mu\nu}\right)\,,
\end{split}
\end{equation*}
where the renormalised boundary EMT is 
\begin{subequations}
    \begin{align}
       T_{\text{ren}}^\mu &= -Mv^\mu + h^{\mu\sigma}v^\rho\left(\partial_\rho\tilde b_\sigma-\partial_\sigma \tilde b_\rho\right)\,,\\
       T_{\text{ren}}^{\mu\nu}      &= -2P_\rho h^{\rho(\mu}v^{\nu)} - Mh^{\mu\nu}\,.
    \end{align}
\end{subequations}
Here, $2M=S^{(0)}-a^2$,
and $\tilde b_\mu=a_\mu+\frac{1}{d}K\tau_\mu$
transforms as $\delta \tilde b_{\mu}=\partial_\mu\Lambda_D$ under Weyl transformations
(for all $d$). It is a remnant of $b_\mu$ in \eqref{eq:btrafo} after our gauge fixing.
The on-shell action is diffeomorphism invariant as the diffeomorphism Ward identity~\eqref{eq:diffWI} corresponds to the equation we obtain by solving the Einstein equations for~\eqref{eq:3Dsol}.
We also see that the Weyl Ward identity~\eqref{eq:non-anomalous-Weyl-WI} is obeyed, so that the on-shell action is Weyl invariant. There is, however, a boost anomaly: the energy current does not satisfy the boost Ward identity~\eqref{eq:non-anomalous-boost-WI}.
Instead,
\begin{equation}
    h^\mu_\rho T^\rho_{\text{ren}} =\mathcal{A}^\mu_{\text{B}}= h^{\mu\sigma}v^\rho\left(\partial_\rho\tilde b_\sigma-\partial_\sigma \tilde b_\rho\right)\,.
\end{equation}
The right-hand side is a Carroll boost anomaly.
It cannot be removed by a local counterterm, and it obeys the Wess--Zumino consistency conditions~\footnote{See~\cite{Jensen:2017tnb} for a classification of anomalies in $(1+1)$-dimensional warped conformal field theories, which also have a Carroll boost symmetry but a different Weyl symmetry structure. In that work, it was found that there is no relation between the Carroll boost anomaly and a diffeomorphism anomaly. We expect the same result to hold for $(1+1)$-dimensional conformal Carroll theories. In fact, the $c_L$ central charge in the BMS$_3$ algebra comes from the difference between left and right central charges when taking the contraction of two copies of Virasoro to BMS$_3$ \cite{Bagchi:2012cy}. This difference is related to the diffeomorphism anomaly and we expect this to still be the case after taking the limit.}. It can be shown that the presence of the Carroll boost anomaly is responsible for the $c_M$ central charge of the asymptotic BMS$_3$ symmetry algebra~\cite{Barnich:2006av,Campoleoni:2022wmf}.

\textbf{Boundary EMT-news complex for $d=2$.}
Here, $E = r^2 \exp(\beta) e$, which means that the right-hand side of~\eqref{eq:varSEH+Sext2} diverges linearly in $r$. To counter this, we need to add an intrinsic counterterm. To construct an intrinsic counterterm, such as a curvature scalar, we need to discuss the choice of bulk connection on the $r=\Lambda$ cut-off hypersurfaces.
When solving the bulk Einstein equations for the variables $S$, $\beta$ and $\Pi_{\mu\nu}$ it is useful to rewrite the Einstein equations in terms of a $(d+1)$-dimensional connection that is associated with the radial cut-off hypersurfaces. Such a connection is~\footnote{This connection is similar to the ``rigging'' connection of~\cite{Mars:1993mj},
see also~\cite{Freidel:2024emv}.}
\begin{equation*}
   C^\rho_{\mu\nu}  =  -U^\rho\left(\partial_{(\mu}+A_{(\mu}\right) V_{\nu)}+\frac{1}{2}\Pi^{\rho\sigma}\left(2\partial_{(\mu} \Pi_{\nu)\sigma}
   -\partial_\sigma \Pi_{\mu\nu}\right)\,.
\end{equation*}
We chose this connection such that it is equal to \eqref{eq:C-connection} at leading order in $1/r$, and it has very similar properties.
We will denote the associated covariant derivative and curvature tensors by $D_\mu$ and $R[C]_{\mu\nu\rho}{}^\sigma$.

Consider the following intrinsic counterterm
\begin{equation}
\label{eq:internal-counterterm}
    S_{\text{int}}=b\int_{r = \Lambda} d^{3}x\,E\, r{R}[C]\,,
\end{equation}
where $R[C]=\Pi^{\mu\rho}R[C]_{\mu\nu\rho}{}^\nu$
and where $b$ is a real number that must be fixed. If we take $b=-1$, we can remove the divergences in \eqref{eq:varSEH+Sext2}, so that we obtain 
\begin{eqnarray*}       &&\delta\left(S_{\text{EH}}+S_{\text{ext}}+S_{\text{ext}}+S_{\text{int}}\right)\Big\vert_{r\to\infty} =\\
&&2\int d^{3}x e\left(T_{\text{ren}}^\mu\delta\tau_\mu+\frac{1}{2}T_{\text{ren}}^{\mu\nu}\delta h_{\mu\nu}+\frac{1}{2}S_{\text{ren}}^{\mu\nu}\delta C_{\mu\nu}\right)\,.
\end{eqnarray*}
We note that, at order $r$, we encounter the term $S_{\text{ren}}^{\mu\nu}\delta h_{\mu\nu}$ where ${\mathcal{T}}^{(3)}{}^{\mu\nu}=S^{\mu\nu}_{\text{ren}}$, which can be shown to be a total derivative.
This is important because the shear is subleading to $h_{\mu\nu}$ in the expansion of $\Pi_{\mu\nu}$, so this had to happen in order to get a response to varying the shear.
We recall that the variation of $h_{\mu\nu}$ is constrained due to~\eqref{eq:K-constraint}.
When computing $T_{\text{ren}}^\mu$ etc.~we obtain an EMT-news complex whose diffeomorphism Ward identity \eqref{eq:diffWI} agrees with the Bondi mass and angular momentum loss equations.
For $d=2$, these arise from $U^\mu R_{\mu\nu}=0$ at order $r^{-2}$.
We thus conclude that the on-shell action is diffeomorphism invariant.

However, $T_{\text{ren}}^\mu$ computed with the above scheme of counterterms does not (yet) obey the Weyl \eqref{eq:non-anomalous-Weyl-WI} or boost Ward identities \eqref{eq:non-anomalous-boost-WI}. We can add finite local intrinsic counterterms to improve the renormalised EMT-news complex such that it obeys the Weyl Ward identity,
whilst the boost Ward identity is obeyed up to an anomaly.
Rather than writing down these finite counterterms in terms of bulk objects, it is often more practical to first compute $T_{\text{ren}}^\mu$ etc.\ explicitly and then to perform improvements.
All relevant improvements are generated by varying the following diffeomorphism-invariant action with respect to the fields $\tau_\mu, h_{\mu\nu}$ and
$C_{\mu\nu}$,
\begin{equation*}
        \int d^3x \, e\left(C^{\mu\nu}\left(a_1 \mathcal{D}_\mu a_\nu+a_2a_\mu a_\nu\right)
        +a_3KC^2+a_4KF^2\right)\,,
\end{equation*}
where $a_1,a_2,a_3,a_4$ are arbitrary constants. 
The structure of these terms is constrained by powers of shear and derivatives.
For an appropriate choice of $a_1,a_2,a_3,a_4$ we obtain improved responses that satisfy the following properties.
First,
the Weyl Ward identity \eqref{eq:non-anomalous-Weyl-WI} is obeyed, 
which means that $h_{\mu\nu}T^{\mu\nu}_{\text{ren,imp}}$ can be obtained from~\eqref{eq:non-anomalous-Weyl-WI}.
(We use the subscript ``$\text{ren,imp}$'' to indicate that the above are improved versions of $T^\mu_{\text{ren}}$ etc.).
Next,
we obtain an anomalous boost Ward identity
\begin{equation}\label{eq:boostanom}
  h_\rho^\mu T_{\text{ren,imp}}^\rho-\left(\mathcal{D}_\rho -a_\rho\right)S_{\text{ren,imp}}^{\rho\mu} = \mathcal{A}^\mu_{\text{B}}\,.
\end{equation}
The boost anomaly and the remaining components of the
of the EMT-news complex
are given by
{\begin{widetext}
    \begin{subequations}
\begin{align}
\begin{split}
    \mathcal{A}_{\text{B}}^\mu & =  \frac{1}{2}h^{\mu\nu}\left(\partial_\nu+2a_\nu\right)\left(S^{(0)}-a^2\right)+h^{\mu\nu}\left(\mathcal{L}_v-\frac{1}{2}K\right)P^{(0)}_{\nu}+\frac{1}{2}v^\rho\left(\partial_\rho\tilde b_\sigma-\partial_\sigma\tilde b_\rho\right)\left(F^{\sigma\mu}-C^{\sigma\mu}\right)\,,
\end{split}
\\
S^{\mu\nu}_{\text{ren,imp}} & =  \frac{1}{2}h^{\mu\rho}h^{\nu\sigma}N_{\rho\sigma}\,,\label{eq:EMTNews1}\\
\tau_\mu T_{\text{ren,imp}}^\mu & =  -v^\rho v^\sigma g^{(1)}_{\rho\sigma}-\frac{1}{16}K\left(F^2-C^2\right)+\mathcal{D}_\rho\left(h^{\rho\sigma}P^{(0)}_\sigma\right)+\frac{1}{4}F^{\rho\sigma}\left(\partial_\rho\tilde b_\sigma-\partial_\sigma\tilde b_\rho\right)\,,\label{eq:d=2-energy-density-Weyl-cov}\\
    \begin{split}
      h_{\rho\mu}\tau_\nu T_{\text{ren,imp}}^{\mu\nu} & =  \frac{3}{2}P^{(1)}_\rho-\frac{3}{32}a_\rho\left(F^2-C^2\right)+\frac{1}{32}h^\sigma_\rho\left(\partial_\sigma+2a_\sigma\right)\left(F^2-C^2\right)-\frac{1}{2}P^{(0)}_\sigma C^\sigma{}_\rho+\frac{5}{2}P^{(0)}_\sigma F^\sigma{}_\rho\\
    &\quad\,-a_\sigma D^\sigma{}_{\rho}+\frac{1}{2}\left(\mathcal{D}_\sigma-a_\sigma\right)\chi^\sigma{}_\rho\,,
    \end{split}
    \\
    \begin{split}
    h^{\langle\rho}_\mu h^{\sigma\rangle}_\nu T_{\text{ren,imp}}^{\mu\nu} & = \frac{1}{2}\left(S^{(0)}-a^2\right)C^{\rho\sigma} -\frac{1}{2}C^{\alpha(\rho}h^{\sigma)\beta}\left(\partial_\alpha\tilde b_\beta-\partial_\beta\tilde b_\alpha\right)+h^{\alpha\langle\rho}h^{\sigma\rangle\beta}\left(\mathcal{D}_\alpha+3a_\alpha\right)P^{(0)}_\beta\\
    &\quad\,+\frac{1}{2}K D^{\rho\sigma}-\frac{1}{4}h^{\mu\rho}h^{\nu\sigma}\mathcal{L}_v\left(F_{\mu\kappa}C^\kappa{}_\nu\right)+\frac{1}{2}h^{\mu\rho}h^{\nu\sigma}\mathcal{L}_v\chi_{\mu\nu}\,.
    \end{split}
\end{align}
    \end{subequations}
\end{widetext}}
\noindent
The boost anomaly $\mathcal{A}_{\text{B}}^\mu$ cannot be removed by local counterterms and satisfies Wess--Zumino consistency. When writing out $\mathcal{A}_{\text{B}}^\mu$, one finds a divergence of the news and terms containing only boundary metric data. Just like the EMT, the news, defined as the response to the shear, is subject to improvements. In the above we considered an improvement that leads to the following Weyl-invariant news tensor
\begin{equation}
N_{\mu\nu} = -\mathcal{L}_v C_{\mu\nu}-\frac{1}{2}K C_{\mu\nu}\,,
\end{equation}
which is spatial and STF. Additionally, recall that the terms involving $\chi^{\mu\nu}$ encode a fundamental  ambiguity~\eqref{eq:t} due to the constraint~\eqref{eq:K-constraint}. This part of the boundary EMT cannot be fixed. We further remind the reader that $\tilde b_\mu=a_\mu+\frac{1}{2}K\tau_\mu$ transforms as a connection under Weyl transformations.
Finally, we have that
\begin{equation}
    -v^\mu v^\nu g^{(1)}_{\mu\nu}=S^{(1)}+2K\beta^{(2)}-2a^\rho P^{(0)}_\rho+C^{\rho\sigma}a_\rho a_\sigma\,.
\end{equation}
The components $\tau_\mu T_{\text{ren,imp}}^\mu$ and $h_{\rho\mu}\tau_\nu T_{\text{ren,imp}}^{\mu\nu}$ are the energy and angular momentum densities.
Note that $h^{\langle\rho}_\mu h^{\sigma\rangle}_\nu T_{\text{ren,imp}}^{\mu\nu}$ is not related to a particular component of the metric in the way that $\tau_\mu T_{\text{ren,imp}}^\mu$ and $h_{\rho\mu}\tau_\nu T_{\text{ren,imp}}^{\mu\nu}$ are. In other words, one cannot solve for a particular component of the $1/r$ expansion of $g_{\mu\nu}$ in terms of $h^{\langle\rho}_\mu h^{\sigma\rangle}_\nu T_{\text{ren,imp}}^{\mu\nu}$. Contracting \eqref{eq:diffWI} once with $v^\nu$ and once with $h^\nu_\rho$,
using \eqref{eq:non-anomalous-Weyl-WI} as well as \eqref{eq:EMTNews1},
we obtain a Carroll-covariant generalisation of the Bondi loss equations
for general boundary Carroll geometry,
{\begin{widetext}
    \begin{subequations}
\begin{align}
    0 & =  -\left(\mathcal{L}_v-\frac{3}{2}K\right)\left(\tau_\mu T^\mu\right)-\frac{1}{4}N^{\rho\sigma}N_{\rho\sigma}+\left(\mathcal{D}_\mu+a_\mu\right)\left(T^\rho h^\mu_\rho\right)\,,\label{eq:diffeoWtimeproj}\\
    \begin{split}
    0 & =  -\left(\mathcal{L}_v-K\right) P_\kappa+ h_{\kappa\sigma}\mathcal{D}_\mu\tilde {T}^{\mu\sigma}+\frac{1}{2}h^\mu_\kappa\left(\partial_\mu+3a_\mu\right)\left(T^{\rho\sigma}h_{\rho\sigma}+\frac{1}{2}N^{\rho\sigma}C_{\rho\sigma}\right) \\
    &\quad\,+\frac{1}{4}h_{\kappa\sigma}\mathcal{D}_\mu\left(N^{\mu\lambda}C_\lambda{}^\sigma-N^{\sigma\lambda}C_\lambda{}^\mu\right)+T^\sigma h^\mu_\sigma F_{\mu\kappa}-\frac{1}{4}N^{\mu\sigma}h^\nu_\kappa\left(\mathcal{D}_\nu+a_\nu\right) C_{\mu\sigma}\,.\label{eq:diffeoWIspatialproj}
    \end{split}
\end{align}
    \end{subequations}
\end{widetext}}
\noindent
We dropped the ``${\text{ren,imp}}$'' subscript, and we defined
\begin{equation}
    P_\kappa=T^{\mu\nu}\tau_\mu h_{\nu\kappa}\,,\qquad\tilde{T}^{\rho\sigma}=T^{\mu\nu}h^{\langle\rho}_\mu h^{\sigma\rangle}_\nu\,,
\end{equation}
and $h^\mu_\rho T^\rho$ is given by \eqref{eq:boostanom} with $S^{\mu\nu}=\frac{1}{2}N^{\mu\nu}$.

To find the relevant expressions for the case of the standard boundary geometry with a sphere and retarded time~$u$, one has to split $x^\mu=(u,x^A)$ and set $\tau_\mu dx^\mu=du$, $v^\mu\partial_\mu=-\partial_u$ and $h_{\mu\nu}dx^\mu dx^\nu=\gamma_{AB}dx^Adx^B$ for a time-independent 2D metric $\gamma_{AB}$, so that $a_\mu=K=F_{\mu\nu}=0$.

\textbf{Outlook and discussion.} 
Many important details regarding the radial solution of the Einstein equations, which were omitted for brevity, will be given in the upcoming publication~\cite{Hartong:2025WIP}. It would be interesting to compute asymptotic charges and their algebra in our Carroll-covariant Bondi--Sachs phase space by using the boundary energy-momentum-news complex derived above, following for example~\cite{Geiller:2024amx}.
Another important question in the context of flat limits of AdS/CFT
would be to construct a Bondi-type phase space in AdS spacetimes which limits to the above, building on~\cite{Compere:2020lrt,Geiller:2022vto,Campoleoni:2023fug}.
Further outstanding questions include a physical interpretation of the free data encoded in the $D_{\mu\nu}$ tensor in~\eqref{eq:D-tensor-properties}.
Finally, it would be interesting to investigate any potential novel contributions to corner terms~\cite{Ciambelli:2022vot}.

\textbf{Acknowledgements}
We are grateful to Luca Ciambelli, Jos\'e Figueroa-O'Farrill, Laurent Freidel, Marc Geiller, Niels Obers, Romain Ruzziconi and C\'eline Zwikel  for useful discussions.
We also thank the participants of the 2025 annual meeting of the Simons Collaboration on Celestial Holography, where this work was presented.
The work of JH and GO is supported by the Royal Society URF Renewal grant 
URF\textbackslash R\textbackslash 221038.
The work of EH is supported by Villum Foundation Experiment project 00050317, ``Exploring the wonderland of Carrollian physics''. The Center of Gravity is a Center of Excellence funded by the Danish National Research Foundation under grant No.~184. The work of VN is supported by a University of Edinburgh School of Mathematics Studentship.

\bibliographystyle{utphys2}
\bibliography{Masterbibliography}
\end{document}